\documentstyle[prb,twocolumn,aps]{revtex}

\tighten

\begin{document}
\title{Tristability in  a non-equilibrium double-quantum-dot in Kondo
regime}
\author{Gustavo. A. Lara $^{1},$Pedro A. Orellana $^{2}$ and Enrique V. Anda $^{3}$}
\address{$^{1}${Departamento de F\'{\i }sica, Universidad de Antofagasta,}\\
Casilla 170, Antofagasta, Chile.\\
$^{2}${Departamento de F\'{\i }sica, Universidad Cat\'{o}lica del Norte,}\\
Casilla 1280,Antofagasta, Chile.\\
$^{3}${Departamento de F\'{\i }sica, P. U. Cat\^{o}lica do Rio de Janeiro,}\\
C.P. 38071-970,Rio de Janeiro, RJ, Brazil.}
\maketitle

\begin{abstract}
\end{abstract}

\begin{abstract}
Electron tunneling through a non-equilibrium double quantum dot in the Kondo
regime is studied. In the region of negative differential resistance, it is
shown that this system possesses a complex response to the applied potential
characterized by a tristable solution for the current. Increasing the
applied potential or reducing the inter-dot coupling, the system goes
through a transition from a coherent inter-dot regime to an incoherent one.
The different nature of the solutions are characterized and it is shown that
the effects of the asymmetry in the dot-lead coupling can be used to control
the region of multistability. The mean-field slave-boson formalism is used
to obtain the solution of the problem.
\end{abstract}

\bigskip

The Kondo effect in quantum-dots ($QDs$) has been extensively studied in the
last years.\cite{goldhaber,cronenwett} The experimental evidence has
confirmed that many of the phenomena that characterize strongly correlated
metals and insulators, as it is the case of the Kondo effect, are present in
$QDs$. The $QDs$ allow to study systematically the quantum-coherence
many-body Kondo state, due to the possibility of continuous tuning of the
relevant parameters governing the properties of this state, in equilibrium
and non-equilibrium situations.

The electron tunneling through double-quantum-dots ( $DQD$) in the Kondo
regime has received much attention in recent years.\cite
{jeong,aguado,georges,busser,aono,izumida,ivanov,orellana,dong} In
comparison with a single quantum-dot, the double-quantum-dot has a richer
physics. For example in $DQDs$ it is possible to study the competition
between the inter-dot antiferromagnetic spin-spin correlation and the
dot-conduction spin-spin correlation present in its ground state and also it
can be investigated the inter-dot coherence effects that results from the
Coulomb interaction. The type of coupling between the $QDs$ determines the
character of the electronic states and the transport properties of the
artificial molecule. In the tunneling regime, the electronics states are
extended across the entire system and can be constructed from a coherent
state based on the bonding or anti-bonding levels of the $QDs$. Some aspects
of the non-equilibrium transport properties of a simplified $DQD$ in the
Kondo regime, constituted by two identical $QDs$ has recently been studied
\cite{orellana}. When the inter-dot coupling is greater than the level
broadening, there is a region of voltage where this system shows a bistable
behavior characterized by two solutions for the current. This behavior
resembles the $J$-$V$ characteristic curve of a double barrier structure in
the accumulation of charge regime and in doped superlattices. \cite
{goldman,prengel,kastrup,aguado2,anda}However, theoretical and experimental
studies have confirmed that a double barrier structure possesses a more
complex behavior than the one predicted by a bistability. The $J$-$V$ curve
of this system has in many cases a Z-shape, indicating the existence of a
tristable solution for the current.\cite{eaves}

\bigskip\ In this work we analyze the new physics derived from the complex
behavior that characterize the electron tunneling through a non-equilibrium $%
DQD$ in the Kondo regime. Although this system is essentially different \
from a double barrier heterostructure, we find that, in certain region of
the parameter space, the $J-V$ characteristic curve of both are similar.
However, the problem we analyze has much richer physics. Increasing the
asymmetry of the coupling of the dots with the leads or reducing the
inter-dot coupling, the system goes through a transition from a coherent
inter-dot regime to an incoherent one. The characteristic curve reflects
this transition \ changing from a Z-shape in the incoherent region to a
loop-shape when the dots behave coherently. The different nature of the
solutions are characterized and it is shown that the effects of the
asymmetry in the dot-lead coupling can be used to control the region of
multistability.

\bigskip We adopt the two-fold degenerate Anderson Hamiltonian in the limit $%
U\rightarrow \infty ,$ that is diagonalized using the mean-field slave-boson
formalism. Although this mean-field approximation has, in certain region of
the parameters that define the system, a non-physical solution it has been
shown that the physics related to the negative differential resistance is
not associated to it\cite{orellana}. The $DQD$ is driven out-of-equilibrium
by means of a dc bias voltage $V$ at zero temperature.The double occupancy
in each dot is forbidden and the inter-dot Coulomb interaction neglected. We
introduce the slave-boson operator $b_{\alpha }^{\dagger }$ that creates an
empty state and a fermion operator $f_{\alpha \sigma }$ that annihilates a
single occupied state with spin $\sigma $. To eliminate the possibility of
double occupancy we impose the constraint $Q_{\alpha }\equiv
\sum\nolimits_{\sigma }f_{\alpha \sigma }^{\dagger }f_{\alpha \sigma
}+b_{\alpha }^{\dagger }b_{\alpha }=1$. The annihilation operator of an
electron in the dot $\alpha $ is $c_{\alpha \sigma }=b_{\alpha }^{\dagger
}f_{\alpha \sigma }$. In the mean field approximation the bosonic operators $%
b_{\alpha }^{\dagger }$ and $b_{\alpha }$ are replaced by their expectation
values, $\left\langle b_{\alpha }\right\rangle =\widetilde{b}_{\alpha }\sqrt{%
2}=\left\langle b_{\alpha }^{\dagger }\right\rangle =\widetilde{b}_{\alpha
}^{\dagger }\sqrt{2}.$ Hence the Hamiltonian of the double quantum-dot
connected to leads plus constraints is written as,

\begin{eqnarray}
H &=&H_{lead}+\sum\limits_{\alpha =0,1,\sigma }\widetilde{\varepsilon }%
_{\alpha }n_{\alpha \sigma }  \nonumber \\
&+&\widetilde{V}_{L}\sum_{\sigma }(c_{-1\sigma }^{\dagger }f_{0\sigma
}+f_{0\sigma }^{\dagger }c_{-1\sigma })+\widetilde{V}_{R}\sum_{\sigma
}(f_{1\sigma }^{\dagger }c_{2\sigma }+c_{2\sigma }^{\dagger }f_{1\sigma })
\nonumber \\
&+&\widetilde{t}_{c}\sum_{\sigma }(f_{0\sigma }^{\dagger }f_{1\sigma
}+f_{1\sigma }^{\dagger }f_{0\sigma })+\sum\limits_{\alpha =0,1}\lambda
_{\alpha }(\widetilde{b}_{\alpha }^{\dagger }\widetilde{b}_{\alpha }-1).
\end{eqnarray}

\noindent where $\widetilde{\varepsilon }_{0}=\varepsilon _{0}+\lambda _{0}$%
, $\widetilde{\varepsilon }_{1}=\varepsilon _{1}+\lambda _{1}$, $\widetilde{V%
}_{L}=V_{L}\widetilde{b}_{0}$, $\widetilde{V}_{R}=V_{R}\widetilde{b}_{1}$, $%
\widetilde{t}_{c}=t_{c}\widetilde{b}_{0}\widetilde{b}_{1},$the $\lambda
_{\alpha}$ are Lagrangian multipliers which guarantee the constraint
conditions on the $Q_{\alpha }$ and the parameters $t_{c},$ $V_{L}$ and $%
V_{R}$ are the inter-dot connection and \ left and right dot connections
with the leads respectively.

\bigskip \noindent The first term on the right-hand side of Eq.1 represents
the electrons in the left and right leads;

\begin{equation}
H_{lead}=\sum_{i\neq 0,1}\varepsilon _{i}n_{i\sigma }+t\sum_{<ij\neq
0,1>\sigma }c_{i\sigma }^{\dagger }c_{j\sigma }.
\end{equation}

\noindent where the operator $c_{i\sigma }^{\dagger }$ creates an electron
in the site $i$ with spin $\sigma $, $\varepsilon _{i}$ is the site energy
and $t$ is first-neighbor hopping in the leads.

Here $H_{lead}$ corresponds to the free-particle Hamiltonian with
eigenfunctions of the Bloch type,

\begin{equation}
\left| k,\sigma \right\rangle =\sum_{j}e^{ikja}\left| j,\sigma \right\rangle
\end{equation}

\noindent where $\left| k,\sigma \right\rangle $ is the momentum eigenstate
with spin $\sigma $ and $\left| j\right\rangle $ is a Wannier state
localized at site $j $ of spin $\sigma $. The dispersions relation
associated with these Bloch states reads

\begin{equation}
\varepsilon _{k}=2t\cos (ka)
\end{equation}

The model has an energy band, extending from $-2t$ to $+2t$, with the first
Brillouin zone expanding the $k$ interval $[-\pi /a,\pi /a]$.

The stationary state of the complete Hamiltonian $H$ with energy $%
\varepsilon _{k}$ can be written as
\begin{equation}
\left| \psi _{k\sigma }\right\rangle =\sum_{j}a_{j\sigma }^{k}\left|
j,\sigma \right\rangle
\end{equation}
\noindent where the coefficients $a_{i\sigma }^{k}$ are given by,

\begin{equation}
a_{j\sigma }^{k}=<j,\sigma |\psi _{k\sigma }>.
\end{equation}

We obtain the following eigenvalues equations for the Wannier amplitudes $%
a_{j\sigma }^{k}$

\begin{eqnarray}
\varepsilon _{k}a_{j,\sigma }^{k} &=&\left\langle j,\sigma \left| H\right|
\psi _{k\sigma }\right\rangle  \nonumber \\
\varepsilon _{k}a_{j,\sigma }^{k} &=&\varepsilon _{j}a_{j,\sigma
}^{k}+t(a_{j-1,\sigma }^{k}+a_{j+1,\sigma }^{k})\;\;\;\;(j\neq -1,0,1,2),
\nonumber \\
\varepsilon _{k}a_{-1(2),\sigma }^{k} &=&\varepsilon _{-1(2)}a_{-2(2),\sigma
}^{k}+\widetilde{V}_{L(R)}a_{0(1),\sigma }^{k}+ta_{-2(3),\sigma }^{k},
\nonumber \\
\varepsilon _{k}a_{0(1),\sigma }^{k} &=&\widetilde{\varepsilon }%
_{0(1)}a_{0(1),\sigma }^{k}+\widetilde{V}_{L(R)}a_{-1(2),\sigma }^{k}+%
\widetilde{t}_{c}a_{1(0),\sigma }^{k},
\end{eqnarray}

\noindent where $a_{j,\sigma }^{k}$ is the amplitude of probability to find
the electron in the site $j$ and state $k$ with spin $\sigma $.

The four parameters ($\widetilde{b}_{0}$, $\widetilde{b}_{1}$, $\lambda _{0}$%
, $\lambda _{1}$) are determined minimizing the expectation values of the
Hamiltonian (7). They satisfy the set of four equations:

\begin{eqnarray}
\widetilde{b}_{0(1)}^{2}+\frac{1}{2}\sum\limits_{k,\sigma }\left|
a_{0(1),\sigma }^{k}\right| ^{2} &=&\frac{1}{2},  \nonumber \\
\frac{\widetilde{V}_{L(R)}}{2}\sum\limits_{k,\sigma }{%
\mathop{\rm Re}%
}(a_{-1(2),\sigma }^{k*}a_{0(1),\sigma }^{k}) &+&  \nonumber \\
\frac{\widetilde{t}_{c}}{2}\sum\limits_{k,\sigma }{%
\mathop{\rm Re}%
}(a_{1(0),\sigma }^{k*}a_{0(1),\sigma }^{k})+\widetilde{\lambda }_{0(1)}%
\widetilde{b}_{L(R)}^{2} &=&0.
\end{eqnarray}

\noindent The sum over the wave vector $k$ cover all the occupied states.
The resulting equations are nonlinear because of the renormalization of the
localized levels in the dots, the interdot coupling tunneling and the
coupling tunneling between the $QD$s and the leads.

The stationary scattering problem is solved by finding the eigenstates with
positive $k$. In order to study the solutions of equations (4) and (5) we
assume a plane wave incident from the left with an amplitude $I$, with a
partial reflection amplitude $R$. The waveform at the right is a simple
plane wave with intensity given by the transmission amplitude $T$. Taking
this to be the solution of the system, we can write,
\begin{eqnarray}
a_{j}^{k} &=&Ie^{ikaj}+Re^{-ika_{j}}\text{, }j<0  \nonumber \\
a_{j}^{k} &=&Te^{ik_{R}aj},\text{ }j>1
\end{eqnarray}

\noindent and those with negative $k$

\begin{eqnarray}
a_{j}^{-k} &=&\widetilde{I}e^{-ikaj}+\widetilde{R}e^{ika_{j}},j>1  \nonumber
\\
a_{j}^{-k} &=&\widetilde{T}e^{-ik_{L}aj},j<0
\end{eqnarray}

These functions describe plane-wave electron approaching the scattering
potential from $j=-\infty $ and $j=+\infty ,$ respectively, with
transmission and reflections amplitudes $T$ and $R$ for $k>0$, and $%
\widetilde{T}$ and $\widetilde{R}$ for $k<0$.

The solution of equations (7) can be obtained through an adequate iteration
from right to left for $k>0$, and from left to right for $k<0$. For a given
transmitted amplitude, the associated reflected and incident amplitudes may
be determined by matching the iterated function to the proper plane wave at
the left lead for $k>0$ and at the right lead for $k<0.$ The transmission
amplitude $t(k)=T/I$ and $\widetilde{t}(k)=\widetilde{T}/\widetilde{I}$ are
obtained from the iterative procedure.

The equations (7) and (8) are nonlinear and require a self-consistent
solution that is obtained using a conjugate gradient algorithm. In this
work, we are interested in analyzing all the stationary solutions for the
current. As $J(V)$and $V(J)$ are multivalued functions, the multiple
solutions for the quantities $\widetilde{b}_{0}$, $\widetilde{b}_{1}$, $%
\lambda _{0}$, $\lambda _{1}$are obtained by fixing, first the external \
potential $V$ and afterwards the current$\ J$ in the equations of motion of
the system. This method permits us to obtain the complete $J$-$V$%
characteristic curve.

Once the amplitudes $a_{j,\sigma}^{k}$ are known, the current is numerically
obtained from,

\begin{equation}
J=\frac{2e}{\hbar }\tilde{V}_{L}\sum\limits_{k,\sigma }{%
\mathop{\rm Im}%
}\{a_{-1,\sigma }^{k*}a_{0,\sigma }^{k}\}.
\end{equation}

We study first a model which consists of two leads equally connected$%
(V_{L}=V_{R}=V_{0})$ to two quantum dots with $\mu _{L}=-V/2$ and $\mu
_{R}=V/2$, $t_{\text{ }}=30\;\Gamma _{0}$, $V_{0}=5.48\;\Gamma _{0}$, $%
\varepsilon _{0}=\varepsilon _{1}=-3.5\;\Gamma _{0}$ (Kondo regime with $%
T_{K}\approx 10^{-3}\;\Gamma _{0}$ with $\Gamma _{0}=2\pi V_{0}^{2}\rho (0)$%
).

The Fig 1 shows the $J$-$V$ characteristic curve for different values of $%
t_{c}$. Within a range of the external bias voltage values, there are three
stationary solutions for the current. We can identify two regimes. One for $%
\Gamma _{0}<t_{c}<t_{c0}$ ,( $t_{c0}$ is a constant to be analyzed below ),
where the $J-V$ characteristic curve has a Z-shape and another one for $%
t_{c}>t_{c0}$ where the curve becomes loop-shaped. An experiment that would
consist in studying this system just by simply continuously modifying the
applied potential $V$ would characterize the behavior of the current as
bistable. At a critical value $Vc_{\uparrow ,}$ there would be a
discontinuous reduction of the current when $V$ is increased and other $%
Vc_{\downarrow }(Vc_{\uparrow }>Vc_{\downarrow })$ where the current would
rise, ($t_{c}<t_{c0}$), or diminish , ($t_{c}>t_{c0}),$ discontinuously when
$V$ is reduced. In this case the interior of the bistability remains
inaccessible to the experiment. However, the internal shape could be
reached, as it is for the case of double barrier heterostructrures,
employing a load line in the experimental measurement that permits to go
along all the points of the $J-V$ characteristic curve.

When $t_{c}>\Gamma _{0}$, the Kondo levels ${\widetilde{\varepsilon }_{0}}$
and ${\widetilde{\varepsilon }_{1}}$are splitted into two molecular states,
corresponding to the bonding and antibonding levels. They are located in, $%
\widetilde{\varepsilon }_{\pm }=\{{(\widetilde{\varepsilon }_{0}+\widetilde{%
\varepsilon }_{1})/2\pm \sqrt{(\widetilde{\varepsilon }_{0}-\widetilde{%
\varepsilon }_{1})^{2}+4\widetilde{t}_{c}^{2}}\}/2}$. In the Fig. 2a and 2b
we plot $\widetilde{\varepsilon }_{\pm }$, as a function of $V,$ for two
values of $t_{c}$. The main features are: a) For $V=0$, the bonding and
antibonding levels satisfies $\widetilde{\varepsilon }_{\pm }$ $=$ $\pm
\widetilde{t}_{c}.$ When the bias $V$ is increased of the order of $T_k$, $%
\widetilde{\varepsilon }_{\pm }$ are kept almost constant and the dots
reacts to the external potential in a coherent way. For higher values of $V$
, $\widetilde{\varepsilon }_{+\text{ }}$and $\widetilde{\varepsilon }_{-%
\text{ }}$approach each other and go through the multistable region. The
right dot is connected to the left reservoir by the mediation of the left
one through an effective interaction $V_{L}^{eff}$ that for energies in the
vicinity of the Fermi level of the order of $T_k$, is approximately given by
$V_{L}^{eff}=$ $\widetilde{t}_{c}/2\pi \widetilde{V}_{L}\rho (0)$. The
effective levels have a clear different behavior whether the right dot is
interacting stronger with the left than with the right $V_{L}^{eff}>%
\widetilde{V}_{R}$ or the contrary, $V_{L}^{eff}<\widetilde{V}_{R}$ . For
the case we are analyzing, $(V_{L}=V_{R}=V_{0})$, this can be controlled
manipulating \ the value of $t_{c}$. Defining $t_{c0}$as the frontier
between these two situation, in the case when $t_{c}<t_{c0},$ increasing the
potential beyond the mutistable region, the coherence between the dots is
lost and $\widetilde{\varepsilon }_{+\text{ }}$and $\widetilde{\varepsilon }%
_{-\text{ }}$converge towards their own chemical potential $\mu _{L(R)}$, as
display in Fig.2a$.$ The interactions of the dots with the leads $\widetilde{%
V}_{L}$ and $\widetilde{V}_{R}$ are equally renormalized as can be seen in
Fig 2c. The system possesses an splitted Kondo peak. One sub-peak with its
weight essentially located at the left dot with energy at the neighborhood
of $\mu _{L}$ and the other located at the right dot near $\mu _{R}.$ For $%
t_{c}>t_{c0}$ by the contrary, due to the stronger interaction between the
dots, they maintain their coherence and result to be partially disconnected
from the right lead. This is reflected in the behavior of the renormalized
interaction of each dot with its respective lead as shown in Fig.2d and in
their localized levels $\widetilde{\varepsilon }_{+\text{ }}$and $\widetilde{%
\varepsilon }_{-\text{ }}$, that both follow $\mu _{L}$ as $V$ increases
presented in Fig2b. The Kondo peak, almost pinned at $\mu _{L},$ is in this
case equally distributed among the two dots.

Next we will discuss the effect of the  asymmetric site energies case,  $%
\varepsilon _0\neq \varepsilon _1$.  We model the asymmetric site
energies by setting $\varepsilon _0=(1-\kappa )e_{0,}\varepsilon
_1=(1+\kappa )e_0,$ where $\kappa $ is the asymmetry parameter and
$e_0=-3.5\;\Gamma _{0}$ . Figs 3 shows the $J$-$V$
characteristics for various values of $\kappa $ at $t_c=1.3\Gamma _0$ and $%
t_c=2.5\Gamma _0$.  It is seen that the region of multistability
is very sensitive to the change of $\kappa $ and in particular it
disappears for sufficiently large values of it. The coherent or
incoherent behavior of the dots discussed above can be controlled
in this case changing the asymmetry parameter $\kappa .$ Fig 4
shows a map with the regions of multistability depending on the
bias voltage $V$ and $\kappa $. It is seen that the width of the
multistability region has lower and upper limits, which depend
upon the asymmetry in the dot-lead coupling. Similar results were
obtained for an asymmetric dot-lead coupling case, $V_L\neq V_R$.

In this paper we have studied the $J$-$V$ characteristic curve of a
double-quantum-dot in the Kondo regime. It exhibits a multistable behavior
that enhances as the interdot coupling tunneling increases. It becomes
Z-shaped or loop-shaped depending on the various coupling tunneling
parameters, which control wether the dots respond coherently or incoherently
to the external applied potential. Moreover, we show that it is possible to
manipulate the multistability in a sample by changing the asymmetry in the
dot-lead coupling.

Work supported by grants Milenio ICM P99-135-F, FONDECYT\ 1990443,
C\'{a}tedra Presidencial en Ciencias, Fundaci\'{o}n Antorchas/Vitae/Andes ,
FINEP and CNPq.

\begin{figure}[tbp]
\caption{$J$-$V$ curve for a) $t_{c}=\Gamma_{0}$, b)$t_{c}=1.3\Gamma_{0}$, c)$%
t_{c}=1.6\Gamma_{0}$ and d) $tc=2.5\Gamma_{0}$}
\end{figure}

\begin{figure}[tbp]
\caption{Figures 2 a) and 2 b) level energies $\epsilon _{\pm }$ vs $V$ at $%
t_{c}=1.3\Gamma _{0}$ $t_{c}=1.6\Gamma _{0}$ respectively. Figures 2 c) and
d) the corresponding renormalization of the dot lead coupling}
\end{figure}

\begin{figure}[tbp]
\caption{$J$-$V$ curve for the asymmetric site energies case at a)
$t_{c}=1.3\Gamma_0$ with $\kappa$=0.005 (solid line), 0.015 (dash
line), 0.04 (dot line) and 0.06(short-dash line), b)$t_{c}=2.5
\Gamma_0$ with $\kappa$=0.04 (solid line), 0.08 (dash line), 0.12
(dot line) and 0.16(short-dash line) }
\end{figure}

\begin{figure}[tbp]
\caption{Map $V$ versus $\kappa$ at a) $t_c=1.1\Gamma_0$, b)$%
t_c=1.2\Gamma_0 $, c)$t_c=1.6\Gamma_0$,d)$t_c=2.5\Gamma_0$ }
\end{figure}

\end{document}